\documentclass[aps,prl,superscriptaddress,reprint,amsmath,amssymb]{revtex4-2}

\usepackage{graphicx}
\begin{document}

\title{Fast matter-antimatter separation via Weibel-induced plasma filamentation}

\author{Oliver Mathiak}
\email{oliver.mathiak@hhu.de}
\affiliation{Institut f\"{u}r Theoretische Physik I, Heinrich-Heine-Universit\"{a}t D\"{u}sseldorf, 40225 D\"{u}sseldorf, Germany}

\author{Lars Reichwein}
\affiliation{Peter Gr\"{u}nberg Institut (PGI-6), Forschungszentrum Jülich, 52425 J\"{u}lich, Germany}
\affiliation{Institut f\"{u}r Theoretische Physik I, Heinrich-Heine-Universit\"{a}t D\"{u}sseldorf, 40225 D\"{u}sseldorf, Germany}

\author{Alexander Pukhov}%
\affiliation{Institut f\"{u}r Theoretische Physik I, Heinrich-Heine-Universit\"{a}t D\"{u}sseldorf, 40225 D\"{u}sseldorf, Germany}

\date{\today}

\begin{abstract}
We study the separation of matter and antimatter driven by the growth of the Weibel instability in a matter-antimatter plasma. The plasma under consideration comprises protons and antiprotons initially at rest, along with a relativistic stream of leptons (electrons and positrons). This stream is maintained by an external force, potentially originating from phenomena such as a photon wind. Our findings reveal the rapid onset of a Weibel-type instability, leading to a distinct separation of matter and antimatter. Results from our particle-in-cell (PIC) simulations are compared with an analytical model based on the linearized magnetohydrodynamics equations.
\end{abstract}

\maketitle

\par \textit{Introduction}---The imbalance between matter and antimatter in the observable Universe remains one of the most profound unanswered questions in modern physics. It is well established that the observable Universe is predominantly composed of matter, with antimatter appearing only in trace amounts \cite{Steigman:1976ev, Steigman_2008, Alcaraz_1999}. This asymmetry is puzzling, as standard mechanisms of matter-antimatter creation, such as pair production, do not inherently favor one over the other.  

One hypothesis suggests that the Universe may contain regions dominated by either matter or antimatter, which could have formed in the early Universe when matter-antimatter plasmas were prevalent \cite{Canetti_2012}. Furthermore, magnetic fields, which are ubiquitous throughout the Universe \cite{Giovannini_2018, doi:10.1142/S0218271804004530, Kronberg_1994}, are believed to have originated during this primordial matter-antimatter plasma era \cite{universe9070309, grayson2023electronpositronplasmabbndampeddynamic}.  

Matter-antimatter plasmas also play a central role in some of the most energetic astrophysical phenomena, such as gamma-ray bursts (GRBs). These bursts are thought to arise from ultra-relativistic winds of matter-antimatter plasma ejected by extreme cosmic objects like black holes, pulsars, and quasars \cite{Znajek, RevModPhys.56.255}. GRBs are extraordinarily luminous, allowing their detection over vast distances—billions of light-years—providing rare insights into the early Universe \cite{Piran:1995cj}.  

Despite their significance, studying such phenomena through telescopes is inherently challenging due to their transient and distant nature. This has spurred growing interest in creating matter-antimatter plasmas in laboratory settings, enabling the study of high-energy astrophysical processes like collisionless shocks, magnetic reconnection, and plasma instabilities \cite{PhysRevAccelBeams.20.043401, Huntington:2013ima, Sarri:2015jyr, PhysRevLett.110.255002, PhysRevLett.113.105003, samsonov2024productionmagneticselfconfinementee}. Advances in high-intensity laser technology have made laboratory astrophysics increasingly feasible and relevant \cite{van_Dishoeck_2019}.  

Although significant progress has been made in generating electron-positron pairs in the laboratory \cite{PhysRevAccelBeams.20.043401, Arrowsmith_2024}, creating a full-fledged matter-antimatter plasma remains a substantial challenge. Plasma behavior requires the plasma size to exceed the Debye length, necessitating a sufficiently large number of pairs.  

In this study, inspired by the conditions of the early Universe, we investigate a matter-antimatter plasma composed of electrons, positrons, protons, and antiprotons. The protons and antiprotons are initially stationary, while the leptons stream collinearly in one direction, driven by high-energy photons exerting a constant, charge-independent force via Compton scattering.  

Using two- and three-dimensional particle-in-cell (PIC) simulations, we demonstrate the filamentation of the plasma due to counter-propagating currents and the exponential growth of density instabilities, specifically a Weibel instability. Additionally, we observe the emergence of a distinct high-energy peak in the lepton energy spectrum, driven by the stopping field. By analyzing particle trajectories, we reveal their complex dynamics and interactions with the filament boundaries.  

Finally, we support our numerical findings with an analytical model describing the initial growth of the instability, derived from the linearization of magnetohydrodynamic (MHD) equations.

\par\textit{Setup}---We consider a matter-antimatter plasma, consisting of protons and antiprotons that are initially at rest and relativistic electrons and positrons that stream colinearly in $z$-direction. The plasma is initially homogeneous and therefore charge neutral. Consequently, the plasma is initially in equilibrium. However, this equilibrium is highly unstable due to the counter-propagating currents of the leptons. In a homogeneous plasma, the currents from the electrons and positrons cancel each other exactly. However, small perturbations of the density will lead to effective currents and a strong repelling force and a perturbation in the charge density. To regain equilibrium, the repelled leptons will pull the oppositely charged hadrons with them, which leads to a separation of matter and antimatter.\\
In this paper, we neglect other effects such as matter-antimatter annihilation: for an ultra-relativistic lepton with energy $\varepsilon \gg m_e c^2$ the cross-section of the annihilation process \cite{Greiner} is 
\begin{align}
    \sigma_\mathrm{ann} = \frac{\alpha^2 \pi \hbar^2}{m_e \varepsilon} \left[ \ln \frac{2\varepsilon}{m_ec^2} -1 \right] + \mathcal{O}\left(\frac{\ln 2 \varepsilon/m_ec^2}{\varepsilon/m_ec^2} \right) \; .
\end{align}
Accordingly, the characteristic time of electron-positron annihilation is 
\begin{align}
    \tau_\mathrm{ann} = \frac{1}{c\sigma n_0} = \frac{m_e \varepsilon}{\alpha^2\pi c \hbar^2 n_0} \left(\ln \frac{2\varepsilon}{m_e c^2} -1 \right)^{-1} \; .
\end{align} 
Annihilation effects become negligible when the characteristic time of the plasma $\tau_p = \omega_p^{-1} = \sqrt{\frac{m_e}{4\pi n_0 e^2}}$ is much smaller than that of the annihilation process.
Because the plasma period and annihilation time depend on the density in a different order, we obtain an upper limit for the plasma density, where this assumption is valid,
\begin{align}
    n_0 \ll \frac{4m_e \varepsilon^2}{\alpha^2\pi c \hbar^3} \left(\ln\frac{2\varepsilon}{m_ec^2}-1 \right)^2 \; .
\end{align}
For a particle energy of $\varepsilon = 100 \text{ MeV} \approx 200 m_ec^2$ this results in an upper density limit of $n_\mathrm{max} \sim 10^{40} \mathrm{cm}^{-3}$.\\
An analogous estimation can be made for the much heavier hadrons, where the calculation of the annihilation cross-section is much more complicated. However, from literature \cite{CARBONELL1997345,ZENONI1999405,BIANCONI2011461,article,Carbonell1993} we can estimate the cross-section for proton-antiproton annihilation to be $\sigma_{p\bar{p}} \sim 0.5 \mathrm{b}$. Using this and the plasma frequency for hadrons gives an upper bound for the hadron density $n_\mathrm{max} \sim 10^{37} \mathrm{cm}^{-3}$. \\
In our setup, we assume that the leptons got accelerated by a flux of streaming photons via Compton scattering. Furthermore, we assume a flux of photons to persist during the simulation that interact with the leptons by exerting a constant, charge independent force.\\
Let us consider a single Compton scattering event of a photon, moving in $z$-direction with energy $\varepsilon = E_\gamma / m_ec^2$ resulting in a scattering angle $\theta$. The change in momentum of the lepton in $z$-direction is then
\begin{align}
    \frac{\Delta p_z}{m_e c} = \varepsilon - \frac{\varepsilon\cos\theta}{1 + \varepsilon(1-\cos\theta)} \; ,
\end{align}
while the transversal momentum change will average out over many scattering events.\\
The effective force resulting from the scattering of photons is 
\begin{align}
    F_\mathrm{CS} = \frac{dp_z}{dt} = n_\gamma c \int d\Omega \Delta p_z  \frac{d\sigma}{d\Omega}\,,
\end{align}
with the photon density $n_\gamma$ and the angular differential cross-section $d\sigma/d\Omega$.\\
The cross-section of the Compton scattering processes is given by
\begin{align}
    \frac{d\sigma}{d\Omega} = \frac{1}{2}r_0^2\frac{\varepsilon'^2}{\varepsilon^2} \left(\frac{\varepsilon'}{\varepsilon} + \frac{\varepsilon}{\varepsilon'}-\sin^2\theta\right)\, ,
\end{align}
with $\varepsilon'$ being the energy of the lepton after the scattering event and $r_0 = e^2/m_ec^2 \approx 2.82 \times 10^{-15}$ m the classical electron radius. For a photon energy of $\varepsilon = 100$ MeV and a photon density of $n_\gamma = 10^{30} \mathrm{cm}^{-3}$ this will result in a force $F_\mathrm{CS} \sim 10^{-5}$ N.

\par \textit{Particle-In-Cell Simulations}---We conduct particle-in-cell (PIC) simulations using the code \textsc{vlpl} \cite{VLPL, Pukhov2016} in order to study the behavior of the proposed matter-antimatter mixture. In the subsequent sections, we first investigate the filamentation dynamics. Later on, examine the peak formation in the tail of the lepton energy spectra.

We simulate the plasma on a square domain with length $2\pi\cdot 100~k_p^{-1}$, where $k_p=\omega_p/c$ and time step $t = 0.005\, \omega_p^{-1}$ for up to $\sim 12000\, \omega_p^{-1}$. The grid step is chosen as $h_x = h_y = 1.5~k_p^{-1}$, with 64 particles per cell for leptons and 32 particles for hadrons, resulting in total to $\sim 3 \times 10^{7}$ numerical particles. 

The leptons have an initial longitudinal momentum of $p_\parallel = 100 m_e c$. Thus, our transverse grid steps do resolve the relativistic plasma skin length $l_s \approx \sqrt{\gamma}k_p^{-1}=10k_p$. The transverse momenta of the leptons follow a normal distribution with mean zero and $\sigma = p_\perp / m_e c = 1, 0.1, 0.01$ for different simulation runs. A constant force $F = 0.1 m_ec\omega_p$, that acts in the $z$-direction is applied to the leptons.

\begin{figure}
    \centering
    \includegraphics[width=1\linewidth]{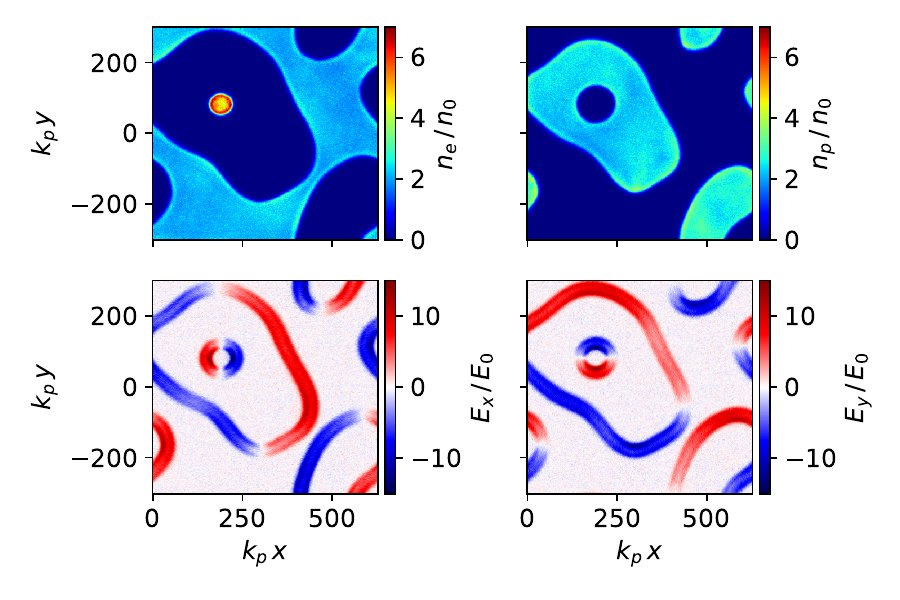}
    \caption{Electron (upper left) and positron (upper right) densities, showing a distinct spatial inversion and the in-plane electric field components (lower) at time $\omega_p t \approx 3000$.}
    \label{fig:enter-label}
\end{figure}

\begin{figure}
    \centering
    \includegraphics[width=1\linewidth]{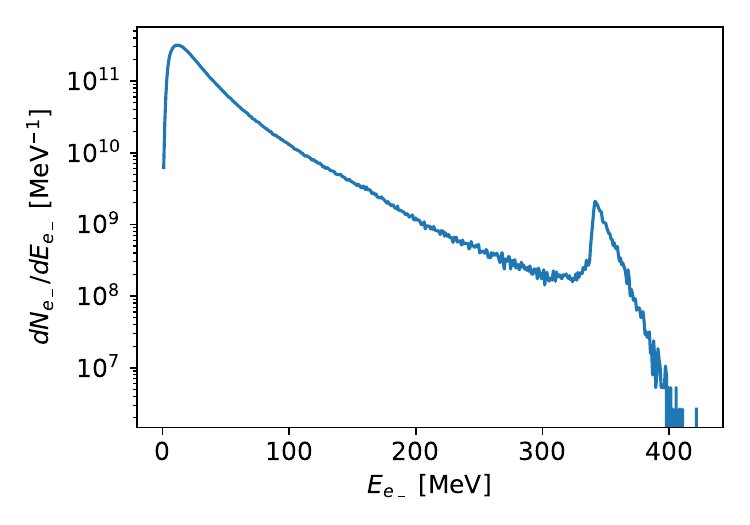}
    \caption{Energy spectrum of electrons at time $\omega_p t \approx 2500$. }
    \label{fig:energy-spectrum}
\end{figure}

\begin{figure}
    \centering
    \includegraphics[width=1\linewidth]{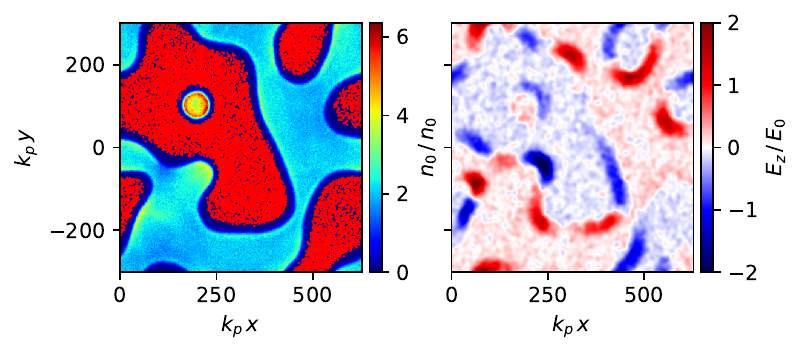}
    \caption{The electron density (left) and the spatial positions of the energy-peak electrons ($E_{e_-} > 300$ MeV) at time $\omega_p t \approx 2500$. Each red dot corresponds to a numerical particle. On the right is the longitudinal electric field after a Gaussian filter has been applied.}
    \label{fig:stopping-field}
\end{figure}

\begin{figure*}
    \centering
    \includegraphics[width=1\linewidth]{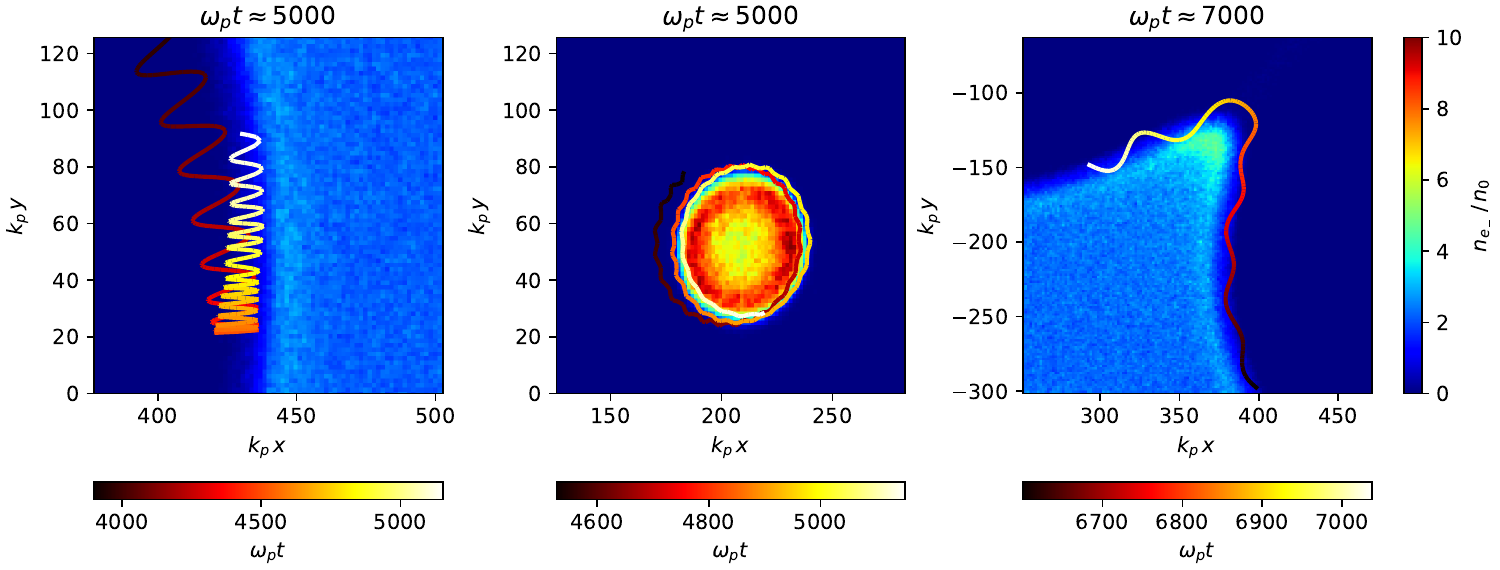}
    \caption{Trajectories of singular electrons along filament edges, showing reflection (left), gyration (middle) and bending along filament edges (right).}
    \label{fig:Trajectories}
\end{figure*}

Due to the isotropic and stationary nature of the problem we use a Yee-lattice based Maxwell solver \cite{Yee} and quadratic momentum pusher. Additionally, to simulate an infinite plasma, we set periodic boundary conditions throughout, both for particles and fields.
\begin{figure*}
    \centering
    \includegraphics[width=1\linewidth]{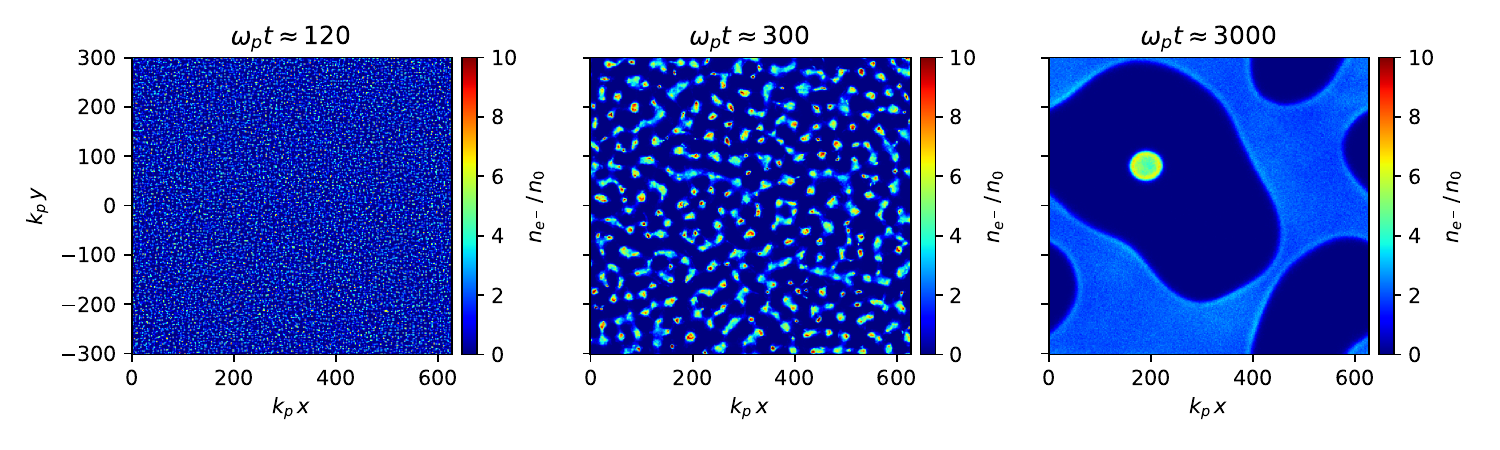}
    \caption{Evolution of the electron density for a 2D simulation. The colorbars are clipped for easier comparison of the various stages.}
    \label{fig:time-evolution}
\end{figure*}
Figure \ref{fig:time-evolution} shows the electron density at different stages of the simulation. At the early stage ($\omega_pt\le 60$), the density-perturbation $\delta n = n - n_0$ is noise-like with $\delta n \ll n_0$. In the intermediate stage ($\omega_pt\le 300$), $\delta n$ grows larger than $n_0$ due to the agglomeration of matter and antimatter in small, distinct patches. In the later stages of the simulation, those patches merge into larger, contiguous filaments that are separated by bands of strong electromagnetic fields (see Fig. \ref{fig:enter-label}).\\
\begin{figure}
    \centering
    \includegraphics[width=\linewidth]{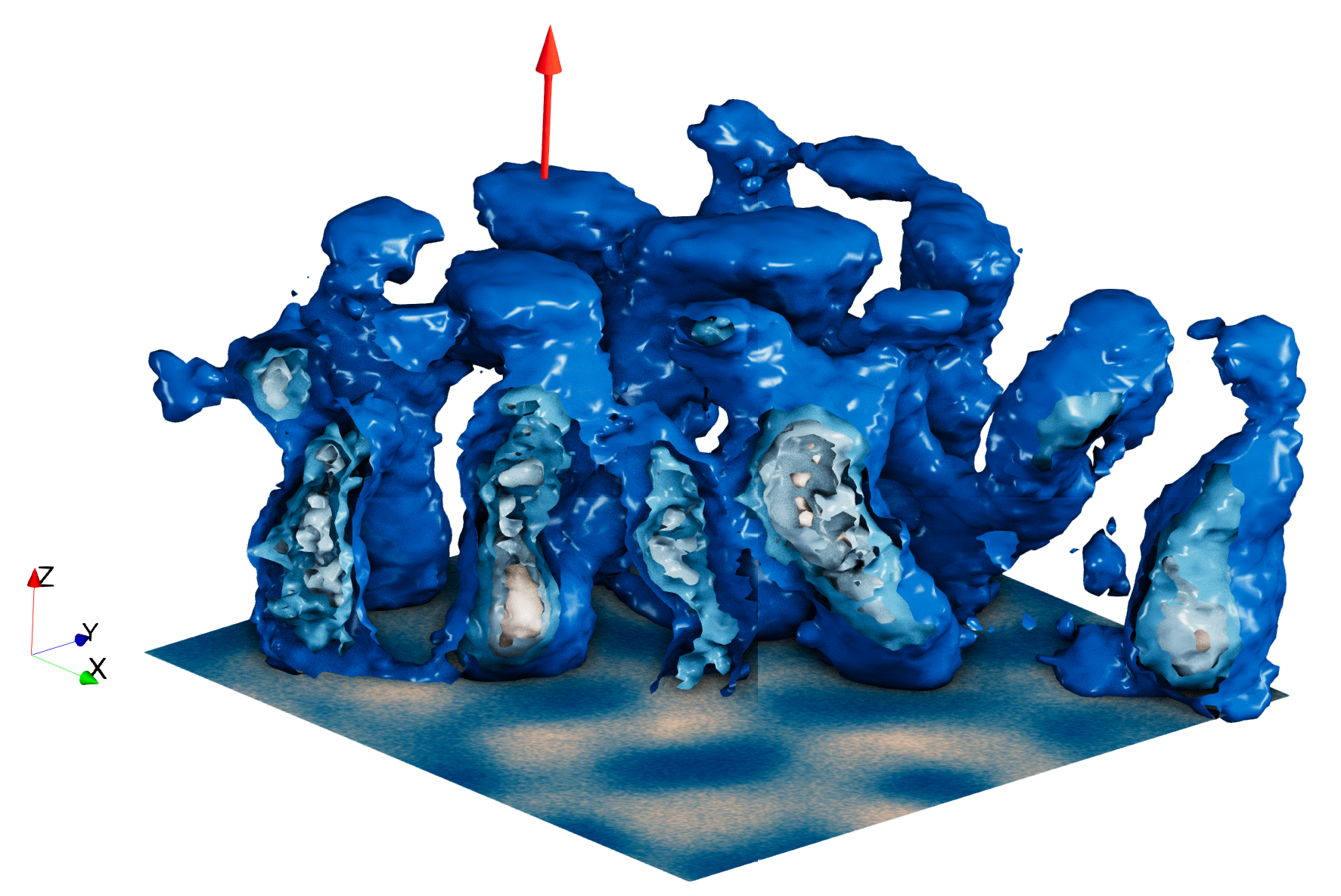}
    \caption{\label{fig:3d} Isosurfaces of electron density for a three-dimensional simulation with $p_\parallel = 100 m_e c$, $p_\perp = 1 m_ec$ at time $\omega_p t \sim 3000$. The bottom plane shows a 2D cut of the density in the $x$-$y$ plane. The red arrow denotes the direction of the external forces.}
\end{figure}
\begin{figure}
    \centering
    \includegraphics[width=\linewidth]{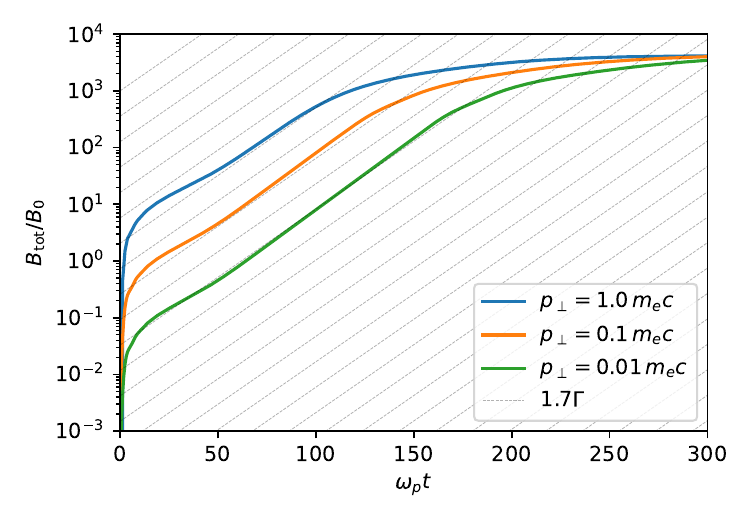}
    \caption{Time evolution of the instability for multiple initial transversal momenta in two dimensions. The gray lines indicate the growth rate in terms of Eq. (\ref{eq:growth-rate}).}
    \label{fig:instability_growth}
\end{figure}
In three dimensions, a similar filamentation occurs with an added structure in $z$-direction. Figure \ref{fig:3d} schematically shows the electron density in three dimensions after $\omega_pt \sim 3000$. An identical filamentation in the $x$-$y$-plane, perpendicular to the direction of momentum, occurs with an added, yet less dominant filamentation in the $z$-direction. This implies that the the two-dimensional case is sufficient to understand the underlying mechanics. Therefore, and for computational efficiency, we will restrict ourselves to 2D-simulations.\\
During the early stage ($\omega_pt \le 100$), the difference in density is still small and can be treated as a perturbation. This is essentially a Weibel instability \cite{Weibel}.\\
It is therefore expected that at the beginning, the instability grows exponentially. This is confirmed for $\omega_pt \le 100$ considering Fig. \ref{fig:instability_growth}. We run the simulation for different initial transverse momenta both in two and three dimensions. While a higher $p_\perp$ initially leads to a faster separation of matter and antimatter resulting in a generally higher $B_{\text{tot}} = \sum_{i,j} |\mathbf{B}|_{i,j}$, the speed at which the instability grows due to the counter-propagating currents is largely unaffected by it.\\
The filamentation of the plasma leads to an interesting characteristic in the energy spectrum of the leptons: whereas the bulk of the particles follow a thermal-like distribution with a maximum at low energies, a distinct peak emerges at the tail of the spectrum, see Fig. \ref{fig:energy-spectrum}.\\
The high-energy electrons, that make up this peak are homogeneously distributed in the ``positron areas'' (cf. Fig. \ref{fig:stopping-field}). In this, area the $z$-component of the electric field that normally counteracts the movement of the leptons, will be accelerating instead of decelerating for particles of the opposite charge, leading to the prominent peak.\\
Similarly, the positron energy spectrum exhibits a peak (not shown here). These particles are, again, located within a filament of the opposite species, i.e. in an electron filament.
While the overall behavior of the plasma can be well understood, individual particle trajectories can be quite complicated, as they interact with the strong fields of the border regions between filaments.
Three examples of such trajectories are given in Fig. \ref{fig:Trajectories}:
(i) electrons may be reflected during their motion along the edge. During this reflection, the frequency of oscillations around the guiding center increases until the electron is finally reflected. This behavior is reminiscent of magnetic bottles, where trapped particles gain increasing transverse momentum (and lose longitudinal momentum) when approaching the bottle's boundary. (ii) Around filaments with a rather homogeneous magnetic field structure, a gyration motion can be observed. The lifetime of this gyration motion depends on how long the respective filament (and the corresponding magnetic field) persists. (iii) Lastly, particles may bent around a filament boundary due to the prevailing fields. The various modes of particle motion make an in-depth analysis rather involved which should therefore be subject of a separate publication.

\par\textit{Analytical Model}---At the early stage, where $\delta n \ll n_0$, we can treat the instability as a small perturbation of a system in equilibrium. Accordingly, we can describe the dynamics of the system using multi-fluid magnetohydrodynamics. \\
For simplicity, we assume an equal initial density for all particle species $n_0 = n_{0e^-} = n_{0e^+} = n_{0p^+} = n_{0p^-}$. Furthermore, we assume that leptons that flow along the $z$-axis initially with the same momentum $\mathbf{p}_{0e^-} = \mathbf{p}_{0e^+} = p_{0} \mathbf{e}_z$ and $|\mathbf{p}_0| \gg mc$, i.e. the particles are ultra-relativistic. This configuration is charge- and current-neutral and therefore in equilibrium. However, as the streams of electrons and positrons correspond to currents of opposite directions, this configuration is unstable. The opposite currents repel each other and start to separate in space transversely to their direction. This separation of leptons pulls hadrons of opposite charge as the charge neutrality must be satisfied. As the electrons and protons must have the same density, we denote it as $n_m = n_{e^-} = n_{p^+}$. Similarly, we denote the antimatter particles as $n_a = e_{e^+} = n_{p^-}$. For simplicity's sake we omit the constant force $F_\mathrm{CS}$ here.\\
The continuity equations are
\begin{equation}
\partial_t n_i+\nabla \cdot\left(n_i \mathbf{v}_i\right)=0 \; , 
\end{equation}
where the index $i$ denotes either sort of particles.
The equations of motion for the leptons
\begin{equation}
\gamma m\left[\partial_t \mathbf{v}_i+\left(\mathbf{v}_i \nabla\right) \mathbf{v}_i\right]=e_i \mathbf{E}-\mathbf{i} \mathbf{k} T \frac{\delta n}{n}+\frac{e_i}{c} \mathbf{v}_i \times \mathbf{B} \; .
\end{equation}
Here, $i=\{e^-, e^+\}, \gamma = \mathrm{const}$ is the relativistic $\gamma$-factor of the leptons, $m$ is their mass, and $T$ is the transverse ``temperature'' of leptons.\\
The equations of motion for hadrons
\begin{align}
    M \partial_t \mathbf{v}_i=e_i \mathbf{E} \; , 
\end{align}
with $i=\{p^-, p^+\}$ and $M$ is the proton mass. The Alfv\'{e}n law is
\begin{align}
    \nabla \times \mathbf{B}=\frac{4 \pi}{c} \mathbf{j} \; .
\end{align}
We make the ansatz
\begin{align}
    n = n_0 + \delta n = n_0 + \delta n_0 \exp(\Gamma t - ikx) \; .
\end{align}
Similarly, we assume the fields and momentum to have an identical, exponentially growing perturbation. Furthermore, we choose the instability wave vector to be in $x$-direction $\mathbf{k} = k \mathbf{e}_x$. The magnetic field has the only component $\mathbf{B} = B \mathbf{e}_y$ and the electric field $\mathbf{E} = E \mathbf{e}_x$.\\
This results in
\begin{equation}
    \Gamma \delta n+n_0 i k v=0 \; ,
\end{equation}
\begin{equation}
    \gamma m \Gamma v=e E- i k T \frac{\delta n}{n}-e B\; ,
\end{equation}
\begin{equation}
    -M \Gamma v=e E \; ,
\end{equation}
\begin{equation}
    i k B=8 \pi e \delta n \; .
\end{equation}

Therefore, we have
\begin{equation}
    (\gamma m+M) \Gamma v=-i k T \frac{\delta n}{n}-e B \; ,
\end{equation}
so that we obtain the dispersion relation for the instability
\begin{align} \label{eq:growth-rate}
    \Gamma = \sqrt{\omega_m^2-k²c_s^2} = \sqrt{\frac{8\pi e^2 n_0}{M+\gamma m} - k^2 \frac{T}{M+\gamma m}} \; , 
\end{align}
where $\omega_m$ is the combined ``matter'' plasma frequency defined by the total mass of the particles and $c_s = \sqrt{T/(M+\gamma m)}$ the sound velocity.\\
The predicted growth rate is in good agreement with the early stages of our simulation ($\omega_p t \le 30$), as is shown in Fig. \ref{fig:instability_growth}. At later stages, it deviates by a factor $\sim 1.7$, indicating a breakdown of our linearization approximation. Additional simulations (not shown here) show that high-frequency perturbations may lead to the additional factor.

\par\textit{Conclusion}--- Using particle-in-cell simulations, we have demonstrated that a system consisting of homogeneously distributed electrons, positrons, protons, and antiprotons can develop counter-propagating currents when exposed to a flux of photons. These currents accelerate the lighter leptons, creating a highly unstable equilibrium. Small density perturbations in this configuration grow exponentially, leading to the filamentation of the matter-antimatter plasma. The resulting structure consists of distinct regions of matter and antimatter, separated by bands of intense electromagnetic fields. The two- and three-dimensional simulation results are in strong agreement.

We observed the emergence of a distinct peak in the energy spectra of both electrons and positrons, originating from leptons confined in the filaments of the opposite species. These leptons are further accelerated by the stopping fields within the filaments. 

Additionally, we studied the complex interactions of individual particles with the fields in the boundary regions between filaments. These interactions include processes such as reflection, gyration, and trajectory bending.

Finally, we compared the simulation results with an analytical model derived from the linearization of magnetohydrodynamics, showing good agreement in describing the early stages of instability growth.

\begin{acknowledgments}
    This work has been supported by the Deutsche Forschungsgemeinschaft and by BMBF (project 05P24PF1).\\
    We gratefully acknowledge the Gauss Centre for Supercomputing e.V. \cite{gauss-centre} for funding this project (lpqed) by providing computing time through the John von Neumann Institute for Computing (NIC) on the GCS Supercomputer JUWELS at Jülich Supercomputing Centre (JSC).
\end{acknowledgments}

\bibliography{bib.bib}

\end{document}